\documentclass[trackchanges]{aastex701}

\begin{document}

\title{Rotational Doppler Cartography of Technosignatures on Unresolved Planets}

\author[orcid=0000-0002-3034-5769,sname='Takahashi']{Keitaro Takahashi}
\affiliation{Kumamoto University}
\email[show]{keitaro@kumamoto-u.ac.jp}  

\begin{abstract}
The discovery of many Earth-like planets has renewed interest in whether life and technological civilizations exist elsewhere. The Search for Extraterrestrial Intelligence (SETI) seeks evidence for technological civilizations via technosignatures across the electromagnetic spectrum. Here, focusing on artificial radio emissions with extremely narrowband signals, we model Earth as a distant, unresolved source and simulate its narrowband transmissions as observed with current and forthcoming radio facilities. Planetary rotation induces small but coherent Doppler drifts (maximum fractional shift of order $10^{-6}$) that imprint a characteristic, time-varying pattern on the spectrum. We develop a forward-inverse framework that exploits this modulation: adopting a population-weighted model for terrestrial transmitters, we compute time-resolved spectra and then apply a new inversion method that reconstructs the underlying transmitter distribution from the temporal pattern of fractional frequency offsets. In noise-added tests, the method recovers the low-order spherical-harmonic structure of the map and retrieves major population centers despite the north-south degeneracy of unresolved observations. The recovered distribution is expected to correlate with continents, climate zones, and population density. This approach moves SETI beyond mere detection, enabling quantitative cartography of a civilization’s activity and inference of host-planet properties through sustained, time-resolved spectroscopy.
\end{abstract}

%% Keywords should appear after the \end{abstract} command. 
%% The AAS Journals now uses Unified Astronomy Thesaurus (UAT) concepts:
%% https://astrothesaurus.org
%% You will be asked to selected these concepts during the submission process
%% but this old "keyword" functionality is maintained in case authors want
%% to include these concepts in their preprints.
%%
%% You can use the \uat command to link your UAT concepts back its source.

%% From the front matter, we move on to the body of the paper.
%% Sections are demarcated by \section and \subsection, respectively.
%% Observe the use of the LaTeX \label
%% command after the \subsection to give a symbolic KEY to the
%% subsection for cross-referencing in a \ref command.
%% You can use LaTeX's \ref and \label commands to keep track of
%% cross-references to sections, equations, tables, and figures.
%% That way, if you change the order of any elements, LaTeX will
%% automatically renumber them.

%%%%%%%%%%%%%%%%%%%%%%%%%%%%%%%%%%%%%%%%%%%%%%%%%%%%%
\section{Introduction\label{sec:introduction}}
%%%%%%%%%%%%%%%%%%%%%%%%%%%%%%%%%%%%%%%%%%%%%%%%%%%%%

Just as the Sun hosts eight orbiting planets, other stars almost certainly harbor planetary systems of their own. Over the past decade, thousands of exoplanets have been discovered, including dozens of rocky, Earth-size planets residing in the circumstellar habitable zone where surface liquid water is plausible \citep{Gillon,Dittmann}. These discoveries have sharpened interest in whether life and, in particular, technological civilizations, exist elsewhere. The Search for Extraterrestrial Intelligence (SETI) seeks evidence for such civilizations via technosignatures across the electromagnetic spectrum, such as radio transmission as well as optical/near-infrared laser emission. In the radio regime, these signals are expected to be exceptionally narrow in bandwidth compared to natural astrophysical emission. Since the pioneering observations of \citet{Drake}, increasingly sensitive facilities operating between 100~MHz and 100~GHz have conducted targeted and wide-field searches for narrowband signals \citep{2016AJ....152..181H,2017ApJ...849..104E,2019AJ....157..122P,2020AJ....160...29S,2020MNRAS.498.5720W,2020AJ....159...86P,Margot,Tao,Mason,Tremblay}. Looking ahead to the 2030s and beyond, next-generation arrays, including SKA1 (Square Kilometre Array~1), SKA2, and the next-generation Very Large Array (ngVLA), will greatly extend the parameter space accessible to SETI, rendering first detection a realistic prospect \citep{2015aska.confE.116S}. In practice, this extension will come from their higher sensitivity and survey speed, wider instantaneous bandwidth with fine spectral resolution, and modern digital backends that enable high-throughput time-frequency searches (e.g., broad drift-rate trials) together with improved localization and RFI discrimination.

Conventional SETI scenarios often consider a single high-power transmitter on a technological planet \citep{2001ARA&A..39..511T,2019ApJ...884...14S,2022ApJ...938....1L}, yet a more representative picture may involve a large population of comparatively weak sources. On Earth, planetary radars are few in number but attain equivalent isotropically radiated powers (EIRP) as high as $10^{13}\,\mathrm{W}$, whereas airport radars, television broadcasters, and radio stations are far less powerful (EIRP $\approx 10^{10}$, $10^{6}$, and $10^{5}\,\mathrm{W}$, respectively) but vastly more numerous \citep{2015aska.confE.116S}. Because even with forthcoming instruments exoplanets will remain unresolved, a distant observer measures the aggregate emission from all transmitters on the hemisphere facing Earth at a given time. As the planet rotates, distinct geographic regions rotate into and out of view, imprinting a modulation at the rotation period on the integrated signal. This situation has been explored in simulations that model terrestrial leakage from television broadcast and mobile-phone base stations and compute the resulting time variability of Earth’s radio flux as seen from afar \citep{Sullivan,Saide}.

When an artificial signal is sufficiently narrow in frequency, the line-of-sight motion of its sources due to planetary rotation induces a measurable Doppler drift \citep{Sullivan}. For Earth’s equatorial speed of $470\,\mathrm{m\,s^{-1}}$, the maximum fractional frequency offset, defined as the Doppler shift divided by the rest-frame frequency, reaches $\sim 1.6 \times 10^{-6}$. Consequently, whenever a transmitter’s intrinsic fractional bandwidth is comparable to or smaller than this scale, rotation-induced spectral modulation should be detectable in principle.

In this work, we develop a forward-inverse framework for rotational Doppler cartography of such narrowband technosignatures, using Earth as an analog of an emitting technological planet hosting an extraterrestrial civilization. We simulate the time-dependent spectral variations generated by the rotation of the emitting planet and adopt a population-weighted model in which the surface density of transmitters scales with human population as a proxy for technological activity. Because many ground-based systems radiate predominantly in near-horizontal directions, a distant observer receives emission chiefly from sources close to the planetary limb. As the planet rotates, the subset of visible transmitters changes continuously, and the Doppler contribution from each source depends on its latitude. Building on these ingredients, we mathematically formulate an inversion method that reconstructs the planetary transmitter distribution from the temporal pattern of fractional frequency offsets in the integrated spectrum. We then place the observer at a distance of several light-years from this Earth-analog emitter and assess what information about the putative civilization and its host planet can be recovered under a plausible detection scenario. We evaluate the extent to which forthcoming high-precision, time-resolved spectroscopy could reveal the spatial distribution of transmitters and show that even unresolved narrowband signals encode a planet-scale “activity map,” enabling a transition in SETI from mere detection to quantitative cartography. Because the transmitter distribution is expected to correlate with population density and, indirectly, with continental and climatic structure, such reconstructions would provide joint constraints on civilization activity and large-scale planetary geography.

The structure of the paper is as follows. Section~\ref{sec:simulation} introduces the forward model and simulation setup for Earth’s transmitter population and the resulting time-resolved spectra. Section~\ref{sec:reconstruction} derives the analytical relation between the spectrogram and the spherical-harmonic coefficients and presents the inversion formalism. Section~\ref{sec:demonstration} evaluates detectability with current and next-generation facilities and demonstrates map reconstruction on noise-added simulations. Section~\ref{sec:summary} discusses implications, limitations, and prospects, and concludes. For completeness and reproducibility, we provide the detailed derivation of the inversion formalism in Appendix \ref{app:reconstruction}.

%%%%%%%%%%%%%%%%%%%%%%%%%%%%%%%%%%%%%%%%%%%%%%%%%%%%%
\section{Simulating Earth as an exoplanet\label{sec:simulation}}
%%%%%%%%%%%%%%%%%%%%%%%%%%%%%%%%%%%%%%%%%%%%%%%%%%%%%

\begin{figure*}[t!]
\plotone{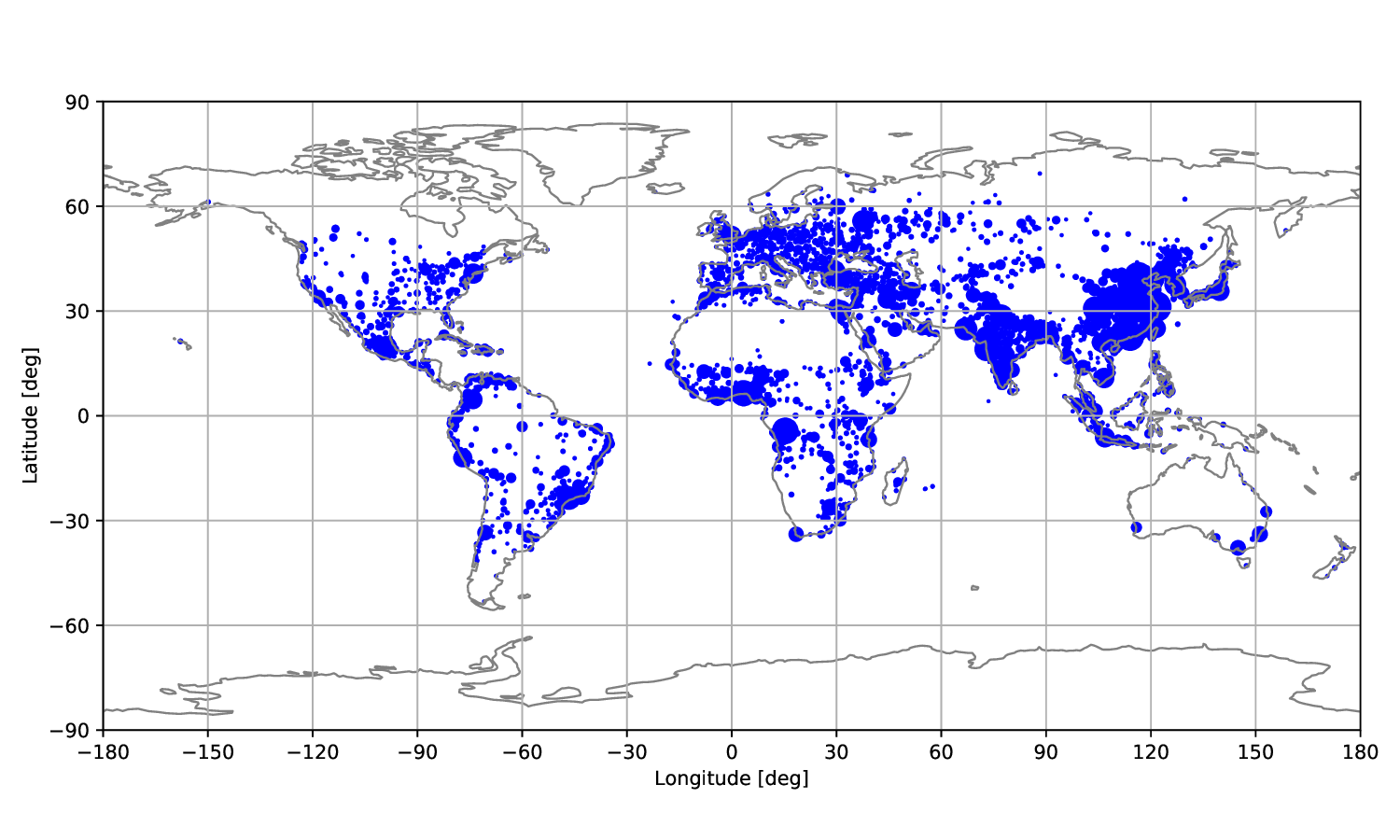}
\caption{Global distribution of cities with populations exceeding 100,000. Point size indicates each city's population, and in our simulation each city's transmitter power is assumed to scale with its population.
\label{fig:population}}
\end{figure*}

\begin{figure*}[t!]
\plotone{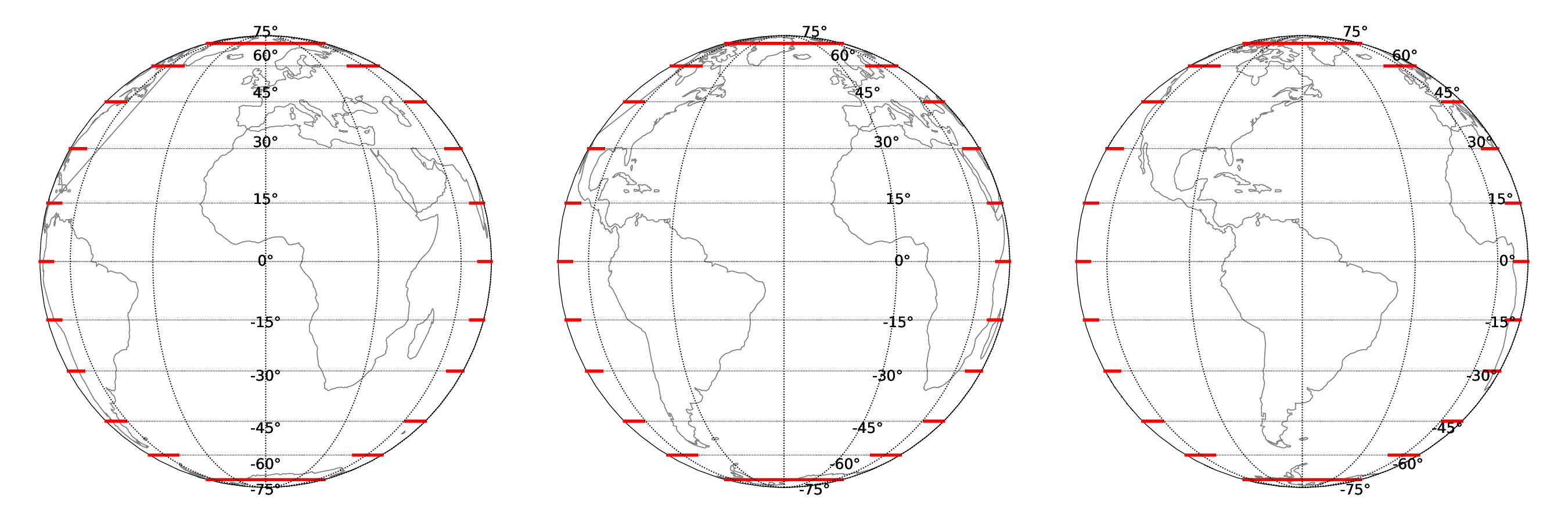}
\caption{Visibility segments on a rotating Earth with coastlines (three central meridians). Each panel shows a circular Earth with coastlines and a latitude-longitude grid; thick red segments near the limb along each latitude mark where a horizontally beamed transmitter would be visible to a distant observer. Panels (A)-(C) are centered on $0^\circ$, $30^\circ$W, and $60^\circ$W, respectively, illustrating how the visibility segments sweep across the world map as the planet rotates eastward. A transmitter moves eastward (left to right) along its latitude and contributes to the signal only while within a red segment. Higher latitudes have longer visible arcs and thus longer dwell times. For clarity, this schematic uses a vertical beam width of $20^\circ$, whereas the simulations use $2.9^\circ$, which yields proportionally shorter arcs.
\label{fig:earth-edge}}
\end{figure*}

In this section, we describe the modeling framework and numerical setup used to simulate Earth as an unresolved exoplanet. We begin by summarizing the forward model and key assumptions that generate time-resolved narrowband spectra from a population of terrestrial transmitters observed at interstellar distances. Fig.~\ref{fig:population} shows all global cities with populations exceeding 100,000. We assign to each city a transmitter whose power scales with its population and assume a common nominal radio frequency. Terrestrial emitters span a wide range of radiation patterns. Here we adopt predominantly near-horizontal emission with an order-of-a-few-degrees vertical beamwidth ($2.9^\circ$ in our simulations), while noting that other classes of transmitters (e.g., scanning or airborne radars) may illuminate higher elevation angles and/or have substantially broader beams. Consequently, only sources located near Earth’s limb are observable from afar (Fig.~\ref{fig:earth-edge}). As Earth rotates, the visible subset of transmitters, and hence the aggregate signal, changes continuously. We include Doppler shifts arising from the rotational motion of each transmitter, noting that equatorial sources exhibit the largest line-of-sight velocities. Although additional Doppler modulation arises from Earth’s orbital motion and that of the observer \citep{2019ApJ...884...14S,2022ApJ...938....1L}, these effects occur on much longer timescales and affect all sources nearly uniformly, leaving the rotation-period signal pattern essentially unaltered after correcting for orbital motion.

To simulate the spectrum of radio emissions from an extraterrestrial civilization, we calculate the leakage of radio power into space from numerous terrestrial transmitters.  We model Earth as a sphere of radius $R$ centered at the origin of a Cartesian coordinate system.  The planet's rotation-axis direction is specified in spherical coordinates, $(\theta, \phi)$, with angular speed $\Omega$.  The observer is assumed to lie far along the positive $x$-axis and can detect signals from any transmitter satisfying $x \le R \gamma$, where $\gamma$ is the transmitter's vertical beam width, taken to be $2.9^\circ$.  When the rotation axis is perpendicular to the observer's line of sight, a single transmitter becomes visible for approximately 12 minutes after rotating into view from the far side.

When the rotation axis points along the $z$-direction $(\theta=0^\circ, \phi=0^\circ)$, a transmitter at latitude $\alpha$ ($-90^\circ \le \alpha \le 90^\circ$) and longitude $\delta$ ($-180^\circ < \delta \le 180^\circ$) has position vector $\vec{r'} = R (\cos\alpha \cos\delta, \cos\alpha \sin\delta, \sin\alpha)$. Here, longitude $\delta$ is measured eastward from an arbitrary prime meridian. We define $\delta=0$ at the beginning of observation ($t=0$) to be the sub-observer (central) meridian, which passes through the point on the surface directly facing the observer. For a general axis orientation ($\theta, \phi$), the position of the source at ($\alpha, \delta$) can be obtained by applying the appropriate rotations to $\vec{r'}$.
\begin{equation}
\vec{r}_0 = R
\left(\begin{array}{c}
\cos\theta \cos\phi \cos\alpha \cos\delta - \sin\phi \cos\alpha \sin\delta + \sin\theta \cos\phi \sin\alpha \\
\cos\theta \sin\phi \cos\alpha \cos\delta + \cos\phi \cos\alpha \sin\delta + \sin\theta \sin\phi \sin\alpha \\
-\sin\theta \cos\alpha \cos\delta + \cos\theta \sin\alpha
\end{array}\right)
\end{equation}
The position vector of a transmitter at time $t$ can be written by replacing $\delta$ with $(\Omega t + \delta)$ as $\vec{r}(t) = \left. \vec{r}_0 \right|_{\delta \rightarrow (\Omega t + \delta)}$, assuming that at $t=0$ the prime meridian ($\delta=0$) lies at the centre of Earth seeing from the observer. Meanwhile, its velocity vector is,
\begin{equation}
\vec{v}(t) = \frac{d\vec{r}(t)}{dt} =
\Omega R \cos\alpha
\left(\begin{array}{c}
- \cos\theta \cos\phi \sin(\Omega t + \delta) - \sin\phi \cos(\Omega t + \delta) \\
- \cos\theta \sin\phi \sin(\Omega t + \delta) + \cos\phi \cos(\Omega t + \delta) \\
\sin\theta \sin(\Omega t + \delta)
\end{array}\right).
\end{equation}
Taking the observer's direction to be $\vec{a}=(1,0,0)$, the line-of-sight component becomes
\begin{equation}
v_{\mathrm{los}}(t)
= \vec{a} \cdot \vec{v}(t)
= - \Omega R \cos\alpha
  \left[ \cos\theta \cos\phi \sin(\Omega t + \delta) + \sin\phi \cos(\Omega t + \delta) \right],
\end{equation}
which determines the significance of Doppler effect. The observed frequency of the signal from the source at ($\alpha, \delta$) is denoted by $f'(\alpha, \delta, t)$, where
\begin{equation}
f'(\alpha, \delta, t) = \left( 1 + \frac{v_{\mathrm{los}}(\alpha, \delta, t)}{c} \right) f_0.
\end{equation}
Here $f_0$ is the rest-frame emission frequency. Hence, the power received by the observer at time $t$ and frequency $f$ can be modeled by the following expression.
\begin{equation}
P(t, f) = \int_{-\pi/2}^{\pi/2} d\alpha \int_{-\pi}^{\pi} d\delta ~ \cos\alpha ~ I(\alpha, \delta)
\frac{1}{\sqrt{2 \pi} \sigma_f} \exp{\left[ - \frac{(f - f'(\alpha, \delta, t))^2}{2 \sigma_f^2} \right]}
\frac{1}{\sqrt{2 \pi} \sigma_b} \exp{\left[ - \frac{x(\alpha, \delta, t)^2}{2 \sigma_b^2} \right]}
\end{equation}
Here, $I(\alpha, \delta)$ denotes the distribution of radio sources on Earth. The forms of the frequency characteristics and beam pattern are assumed to have a Gaussian shape and $\sigma_f$ and $\sigma_b$ are the bandwidth and beam width of the radio emissions, respectively. Using these equations, one can compute the radio spectrum received by the observer from Earth and thereby simulate the spectrogram.

\begin{figure*}[t!]
\plotone{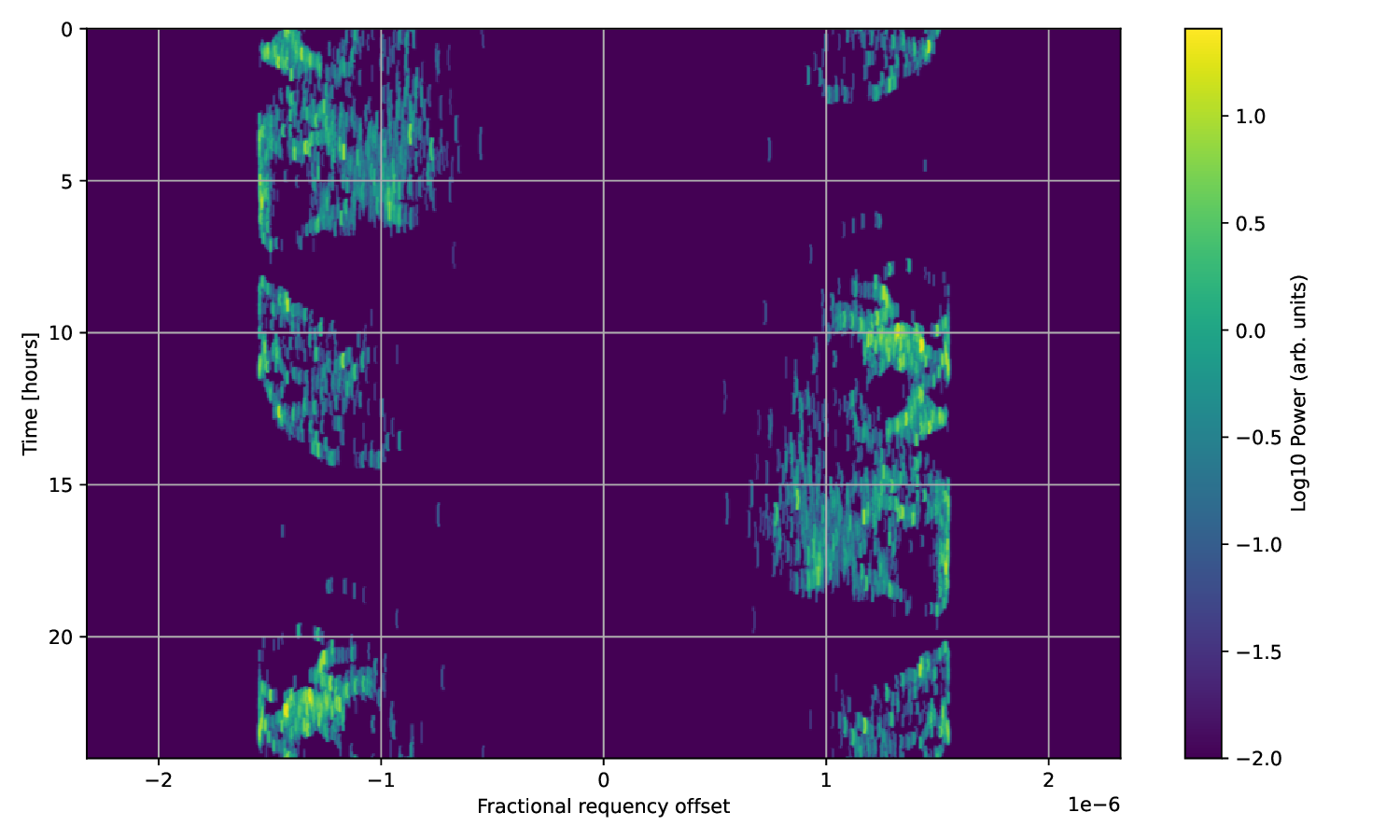}
\caption{Spectrogram when Earth’s rotation axis is perpendicular to the line of sight. (Color scale is logarithmic.) Each pixel is constructed from the sum of populations of cities that are visible to the observer at that time and fractional frequency offset, divided by $10^6$, and plotted as $\log_{10}$ of that quantity. We assume that the corresponding pre-log value is proportional to received signal strength, but its absolute normalization is arbitrary (hence the “arbitrary units” color bar). At time $t=0$, Earth’s prime meridian faces the observer; transmitters at $90^\circ$ W (central North America) produce a positive frequency offset, whereas those at $90^\circ$ E (India) produce a negative offset. Focusing on the positive-frequency side, a bright signal from the U.S. west coast appears, followed by roughly five hours of minimal emission. Around 08:00, transmissions from Japan, Australia, China, and Southeast Asia brighten the signal again, and the bright cluster near 13:00 corresponds to India. As rotation continues, sources in West Asia, Europe, and Africa come into view. Between $\sim$ 17:00 and 19:00, a gap appears at a fractional frequency offset of about $1.4 \times 10^{-6}$, reflecting low population density in three major deserts (the Sahara, Namib, and Kalahari) around $20^\circ$ latitude. By $\sim$ 20:00, South America (followed by North America) rotates into view after the Atlantic. Thus, despite a north–south ambiguity, the spectrogram captures the planet’s transmitter distribution, which is assumed here to follow population.
\label{fig:spectrogram}}
\end{figure*}

First, we consider a simplified case in which Earth's rotation axis is perpendicular to the observer's line of sight as in Fig. \ref{fig:earth-edge}.  Fig. \ref{fig:spectrogram} shows the spectrogram, i.e., the time variation of the radio spectrum.  The vertical axis spans 24 hours, and over such short timescales, during which the orbital phase changes negligibly, this behavior repeats with a 24 h period. The horizontal axis is the fractional frequency offset and low-latitude sources exhibit larger rotational velocities and hence larger frequency offsets, reaching up to $1.6 \times 10^{-6}$ at the equator. For reference, at $f_0 = 1$ GHz a fractional offset of $10^{-6}$ corresponds to an absolute shift of $\Delta f = 1$ kHz, which is readily resolved with Hz-10 Hz channelization commonly used in narrowband SETI analyses. When a transmitter rotates into view from the far side of Earth, it appears with positive offset (blueshifted), and as it rotates out of view toward the far side, it appears with negative offset (redshifted).  Thus, at any given time slice of the spectrum, the positive-offset side contains contributions from the cohort of transmitters located along a particular meridian appearing from Earth's far side, whereas the negative-offset side reflects those on the antipodal meridian.  Because different latitudes correspond to different offsets, the brightness pattern over the range $(0.8 - 1.5)\times10^{-6}$ maps the distribution of sources along those meridians: sources at lower latitudes appear toward the outer edges of this range.  Note that the frequency-offset pattern is symmetric between the northern and southern hemispheres, so that sources at equal northern and southern latitudes overlap in the spectrogram.  Focusing on any single transmitter, it first appears as a blueshifted feature when rotating into view from the far side and then, approximately 12 hours later, reappears as a redshifted feature as it rotates out of view behind the far side.  Consequently, Fig. \ref{fig:spectrogram} exhibits identical patterns on the positive and negative sides, offset by 12 hours.  A more detailed correspondence between geographic source locations and features in the spectrogram is described in the caption of Fig. \ref{fig:spectrogram}.

\begin{figure*}[t!]
\begin{center}
\includegraphics[width=6cm]{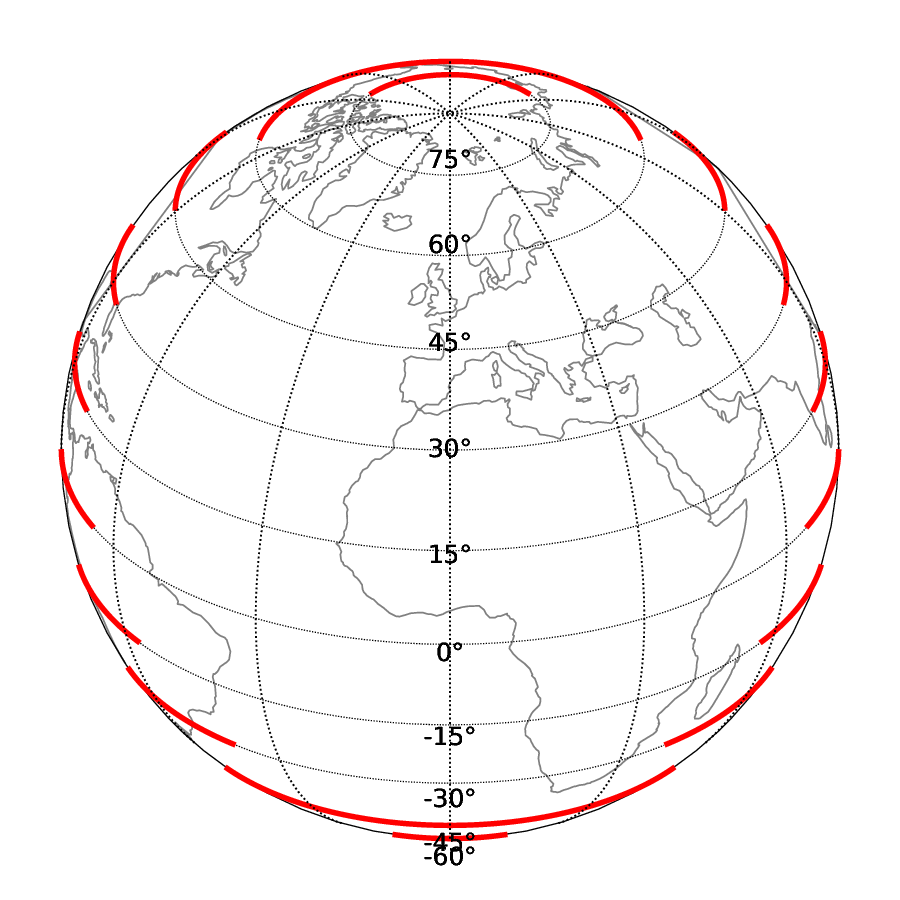} 
\includegraphics[width=10cm]{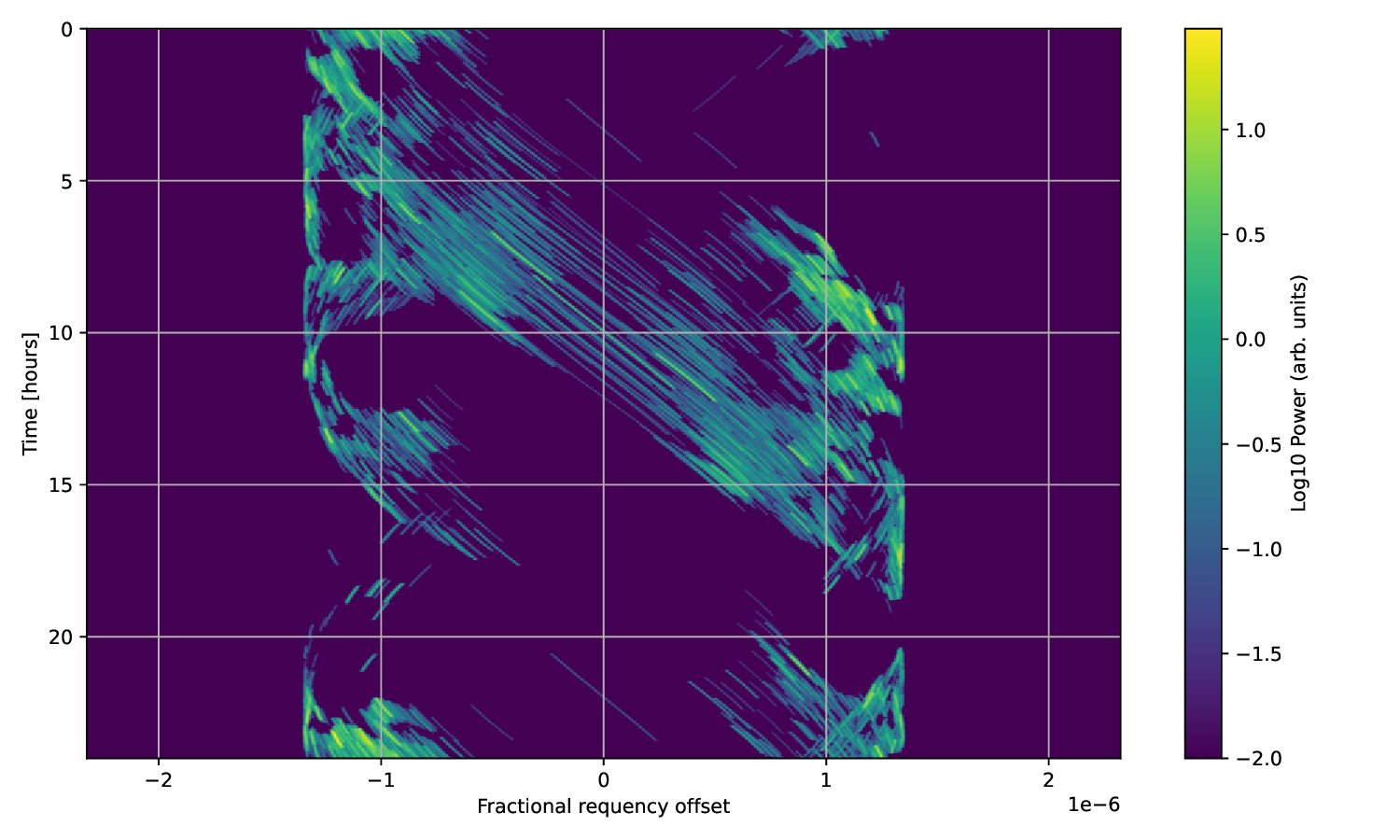} 
\end{center}
\caption{Case with Earth's rotation axis tilted by $30^\circ$. Here Earth's axis is tilted by 30° toward the observer (northward tilt), rather than being perfectly perpendicular. (Left) Map of Earth's latitude/longitude lines and visibility segments as in Fig. \ref{fig:earth-edge}. In practice, only transmitters between about $60^\circ$ S and $60^\circ$ N are observable, while visibility segments also appear at higher latitudes in this figure because for visual clarity this schematic adopts a vertical beam width of $20^\circ$. (Right) Corresponding spectrogram analogous to Fig. \ref{fig:spectrogram} for this tilted orientation.
\label{fig:spectrogram_30}}
\end{figure*}

Next, we consider the case in which the Earth's rotation axis is inclined by $30^\circ$ toward the observer, as shown in the left panel of Fig.~\ref{fig:spectrogram_30}, rather than being perpendicular.  In this configuration, all regions poleward of $60^\circ$ S remain perpetually hidden on the far side of Earth, and thus are never observed.  Conversely, regions poleward of $60^\circ$ N always face the observer, but because the observer's elevation angle as seen from those sources is high, their predominantly horizontal emissions do not reach the observer. Consequently, for an axis tilt of $30^\circ$, the purely geometric near-hemisphere visibility corresponds to latitudes roughly between $-60^\circ$ and $+60^\circ$. Finite vertical beamwidth $\gamma$ broadens this boundary by $\mathcal{O}(\gamma)$.

The right panel of Fig.~\ref{fig:spectrogram_30} shows the resulting spectrogram, which differs markedly from Fig. \ref{fig:spectrogram}.  One set of tracks runs from the upper left to the lower right, while a fainter set runs from the upper right to the lower left.  The former begins at redshift, shifts toward zero offset, and then transitions to blueshift.  This behavior arises from transmitters near $60^\circ$ N: as illustrated in the left panel of Fig. \ref{fig:spectrogram_30}, these sources dwell near the Earth limb for extended durations.  From the perspective of such a high-latitude transmitter, the observer remains just above the horizon for a long time, which is analogous to the prolonged twilight of the midnight sun in high-latitude regions.  Because few major cities lie in this latitude band, these tracks are relatively dim.  The latter, weaker tracks, running from upper right to lower left, originate from sources near $60^\circ$ S. However, low population density in that band renders them largely inconspicuous.

%%%%%%%%%%%%%%%%%%%%%%%%%%%%%%%%%%%%%%%%%%%%%%%%%%%%%
\section{Reconstructing the distribution of radio transmitters\label{sec:reconstruction}}
%%%%%%%%%%%%%%%%%%%%%%%%%%%%%%%%%%%%%%%%%%%%%%%%%%%%%

In this section, we tackle the inverse problem: reconstructing the underlying transmitter distribution from an observed spectrogram such as Fig.~\ref{fig:spectrogram} or the right panel of Fig.~\ref{fig:spectrogram_30}. For readability, we present the main results and a practical reconstruction recipe in the main text, while the step-by-step derivations are given in Appendix~\ref{app:reconstruction}. We first establish an analytical relation between the spectrogram and the spherical-harmonic coefficients $a_{\ell m}$ of the transmitter distribution, and then obtain an explicit inversion for the subset of modes accessible to unresolved observations. For simplicity, we adopt a delta-function model in which both the frequency characteristics and the vertical beam pattern are approximated as delta functions, which yields closed-form expressions that clarify the information content of $P(t, f)$.

By setting the maximum rotational Doppler shift amplitude as $B \equiv f_0 R \Omega / c$ and $\mu = \cos(\pi/2 - \alpha)$, and changing the variable from the frequency itself to the frequency offset ($\Delta f \equiv f - f_0$), we have
\begin{equation}
P(t, \Delta f)
= \int_{-1}^1 d\mu \int_0^{2\pi} d\delta ~ I(\mu, \delta)
~ \delta(\Delta f + B \sqrt{1- \mu^2} \sin(\Omega t + \delta))
~ \delta(\sqrt{1- \mu^2} \cos(\Omega t + \delta)),
\label{eq:Ptf}
\end{equation}
where the first and second delta functions correspond to the frequency characteristics and the beam pattern, respectively. On the other hand, the distribution of radio sources $I(\mu, \delta)$ can be expanded into spherical harmonics.
\begin{equation}
I(\alpha, \delta) = \sum_\ell \sum_{m = - \ell}^{\ell} a_{\ell m} Y_{\ell m}(\pi/2 - \alpha, \delta),
\label{eq:expansion}
\end{equation}
where,
\begin{equation}
Y_{\ell m}(\vartheta, \delta)
= \sqrt{\frac{2 \ell + 1}{4 \pi}} \sqrt{\frac{(\ell - m)!}{(\ell + m)!}} P^m_\ell (\cos\vartheta) e^{i m \delta}
\equiv A_{\ell m} P^m_\ell (\cos\vartheta) e^{i m \delta},
\end{equation}
and $P_\ell^m$ is the associated Legendre function. By substituting this expansion into Eq. (\ref{eq:Ptf}) and performing the integration, we obtain
\begin{equation}
P(t, \Delta f)
= \sum_\ell \sum_{m = -\ell}^{\ell} a_{\ell m} A_{\ell m} e^{-i m \Omega t} (- \mathrm{sgn}(\Delta f) i)^m  \frac{P_\ell^m(\sqrt{1- (\Delta f/B)^2}) + P_\ell^m(-\sqrt{1- (\Delta f/B)^2})}{B \sqrt{1- (\Delta f/B)^2}},
\end{equation}
where $\mathrm{sgn}(\Delta f)$ is the sign function of $\Delta f$ (positive for blueshift and negative for redshift). This is a relation between observable quantity $P(t, \Delta f)$ and information on transmitter distribution $a_{\ell m}$. If we have an inverse relation, which expresses $a_{\ell m}$ as a function of $P(t, \Delta f)$, the distribution map can be reconstructed from the observable quantity $P(t, \Delta f)$.

To do this, first, we perform Fourier transform with respect to $t$.
\begin{eqnarray}
\tilde{P}(\omega, \Delta f)
&=& \int_{-\infty}^\infty P(t, \Delta f) e^{i \omega t} dt \nonumber \\
&=& \sum_{\ell'} \sum_{m' = -\ell'}^{\ell'} a_{\ell' m'} A_{\ell' m'} \delta(\omega - m' \Omega) (- \mathrm{sgn}(\Delta f) i)^{m'}
    \frac{P_{\ell'}^{m'}(\sqrt{1- (\Delta f/B)^2}) + P_{\ell'}^{m'}(-\sqrt{1- (\Delta f/B)^2})}{B \sqrt{1- (\Delta f/B)^2}}
\label{eq:tildeP}
\end{eqnarray}
Then, we evaluate the temporal Fourier transform at harmonics of the emitting planet’s rotation frequency, $\omega = m \Omega ~ (m = 1, 2, 3, \ldots)$. In practice, $\Omega$ can be obtained from the periodicity of the dynamic spectrum itself (and/or from independent rotation constraints such as photometric modulation). By multiplying $\tilde{P}(m \Omega, \Delta f)$ by the associated Legendre functions with a weight of $\Delta f$ and integrating it over $\Delta f$ from $\Delta f = 0$ to $B$, we define $I_{\ell m}$ as,
\begin{equation}
I_{\ell m} = \int_0^B \tilde{P}(m \Omega, \Delta f) \left[ P_\ell^m(\sqrt{1- (\Delta f/B)^2}) + P_\ell^m(-\sqrt{1- (\Delta f/B)^2}) \right] \Delta f ~ d\Delta f,
\label{eq:I_lm}
\end{equation}
which is a quantity calculable from the observable quantity $P(t, \Delta f)$. Substituting Eq.~(\ref{eq:tildeP}), we obtain,
\begin{equation}
I_{\ell m} = a_{\ell m} A_{\ell m} (-i)^m B \left(1+(-1)^{\ell+m}\right) \frac{2 (\ell + m)!}{(2 \ell + 1) (\ell - m)!}.
\end{equation}
We see, when $(\ell + m)$ is even, the expansion coefficients, $a_{\ell m}$, are directly obtainable from the observables, $I_{\ell m}$:
\begin{equation}
a_{\ell m} = \frac{i^m}{2 B} \sqrt{\pi (2 \ell + 1)} \sqrt{\frac{(\ell - m)!}{(\ell + m)!}} I_{\ell m}.
\label{eq:alm}
\end{equation}
Contrastingly, when $(\ell + m)$ is odd, the coefficients $a_{\ell m}$ cannot be obtained from observation data. Therefore, even with noise-free observations the available information is incomplete, so the radio-source distribution cannot be fully reconstructed.  This limitation arises because the signals from the northern and southern hemispheres appear identically and cannot be distinguished observationally.  Consequently, the radio-source map reconstructed from the retrieved $a_{\ell m}$ has the northern- and southern-hemisphere sources superimposed, yielding a distribution symmetric about the equator. Here, it should be noted that we are assuming the value of $B$, the maximum frequency offset, is known in advance from the spectrogram or other observations. In practice, the Doppler span can be estimated from the dynamic spectrum itself as the maximum frequency shift at which emission is detected.

\begin{figure*}[t!]
\plotone{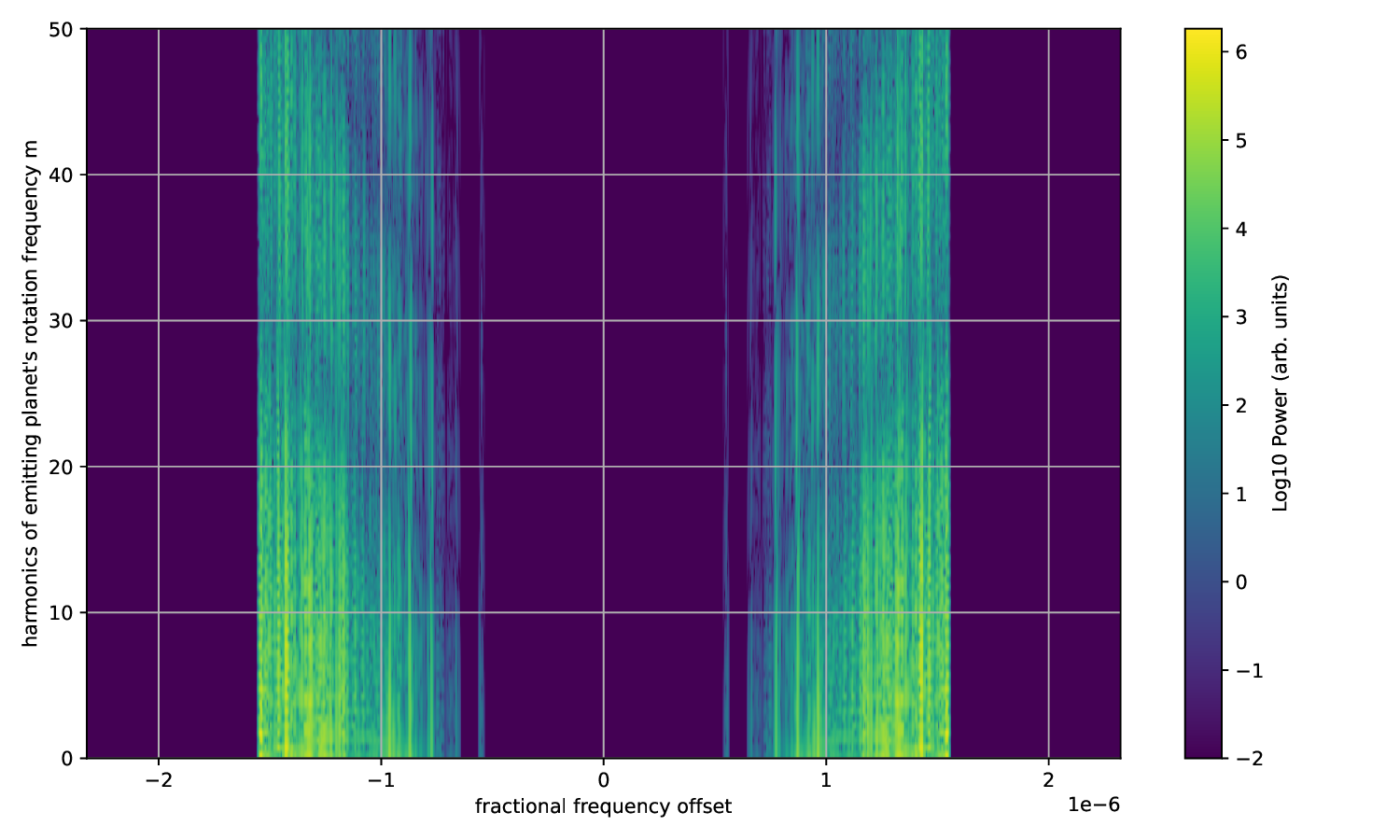} 
\caption{Power spectra, $|\tilde{P(m \Omega, \Delta f)}|^2$. The horizontal axis is fractional frequency offset $\Delta f/f_0$ which corresponds to the latitude of radio sources while the vertical axis is frequency which is conjugate to time and equivalent to $m$-th harmonics of emitting planet’s rotation frequency.
\label{fig:fourier}}
\end{figure*}

Fig. \ref{fig:fourier} displays the power spectrum of the spectrogram shown in Fig.~\ref{fig:spectrogram}, $|\tilde{P}(m \Omega, \Delta f)|^2$, revealing the spatial structure of source distributions along each latitude line specified by $\Delta f$. Lower values of $m$ correspond to large-scale features, while higher values of $m$ capture finer details. Integrating this graph along the horizontal axis with the weight directly relates to the spherical-harmonic expansion coefficients, $a_{\ell m}$, as in Eq. (\ref{eq:alm}).

%%%%%%%%%%%%%%%%%%%%%%%%%%%%%%%%%%%%%%%%%%%%%%%%%%%%%
\section{Detectability and reconstruction performance\label{sec:demonstration}}
%%%%%%%%%%%%%%%%%%%%%%%%%%%%%%%%%%%%%%%%%%%%%%%%%%%%%

In this section, we demonstrate how faithfully the distribution can be recovered using simulations described in the previous section. As a first step, we estimate the detectability of an Earth-analog civilization around the nearest stars using current and next-generation radio telescopes.  Denoting the frequency resolution by $\Delta \nu$, the time resolution by $\Delta t$, and the achievable signal-to-noise ratio by SNR, the maximum distance $d$ at which a transmitter could be detected is estimated as follows \citep{2025AJ....169..118S}.
\begin{eqnarray}
d &=& \sqrt{ \frac{\rm EIRP}{4 \pi ({\rm SNR}) ({\rm SEFD})} \sqrt{ \frac{\Delta t}{\Delta \nu}}}  \nonumber \\
&\approx& 38.0~{\rm ly} \left( \frac{\rm EIRP}{10^{10}~{\rm W}} \right)^{1/2} \left( \frac{\rm SEFD}{1.7~{\rm Jy}} \right)^{-1/2} \left( \frac{\rm SNR}{10} \right)^{-1/2}  \left( \frac{\Delta t}{300~{\rm sec}} \right)^{1/4}  \left( \frac{\Delta \nu}{10~{\rm Hz}} \right)^{-1/4}  
\label{eq:d_max}
\end{eqnarray}
Here, for the EIRP of the transmitter, we adopt a typical value for a high-power, long-range airport radar.  The SEFD (System Equivalent Flux Density) quantifies a radio telescope's sensitivity and we use the SKA1-Mid value of 1.7 Jy at 1 GHz \citep{Braun}.  We assume a frequency and time resolution of $\Delta \nu$ = 10 Hz and $\Delta t$ = 300 sec, respectively.  Inserting these representative numbers shows that for planets around about 500 nearest stars within 38 light years, one could detect a transmitter as bright as a high-power, long-range airport radar at SNR = 10 or better.  To achieve higher SNR or to probe more distant planets, one may employ even more sensitive telescopes.  Furthermore, because the signal is periodic over 24 h, co-adding observations, after correcting for long-term orbital Doppler drifts, can further boost SNR.

To provide a concrete detectability scale, we translate the SNR of a representative time-frequency pixel (bin) in the dynamic spectrum (Fig. \ref{fig:spectrogram} and the right panel of Fig. \ref{fig:spectrogram_30}) into an equivalent EIRP using Eq. (\ref{eq:d_max}). In the simulated spectrograms, the dynamic spectrum is rendered on a 1200 (time) $\times$ 600 (frequency) grid over one 24-hour rotation period, corresponding to a time bin of 72 sec. For additional intuition in the Earth-analog mapping used in the figures, one surface pixel corresponds to 33 km in longitude and 50 km in latitude (area $= 1650 {\rm km}^2$). With the beamwidth of $2.9^{\circ}$, the instantaneous visible limb segment spans 640 km in longitude. Thus, the value plotted at each time-frequency pixel should be interpreted as the summed emission from a limb strip of order (640 km) $\times$ (50 km) = 32,000 ${\rm km}^2$.

We set the noise level such that the brightest pixel in Fig. \ref{fig:spectrogram} has SNR = 5. Although Fig. \ref{fig:spectrogram} does not assume a specific terrestrial transmitter class (nor that terrestrial systems occupy such narrow spectral bins), Eq. (\ref{eq:d_max}) allows a translation from this fiducial SNR to an equivalent EIRP once representative observational parameters are specified. For example, adopting SKA1 SEFD = 1.7 Jy, a distance $d = 10$ light-years, $\Delta t = 72$ sec and $\Delta \nu = 10$ Hz yields an equivalent brightest-pixel EIRP of $\approx 7 \times 10^8$ W. This number should be interpreted as the effective narrowband EIRP contained within that adopted time-frequency bin, summed over the above beam and frequency width. For intuition only, we compare this EIRP scale to familiar terrestrial transmitter classes using the representative values quoted in Section \ref{sec:introduction}. While such systems (e.g., TV broadcast transmitters and airport radars) do not in general emit within the extremely narrow spectral channels considered here, the value EIRP $7 \times 10^8$ W is comparable in scale to the aggregate of about 700 TV-broadcast-class transmitters or about $7\%$ of a typical airport radar.

\begin{figure*}[t!]
\begin{center}
\includegraphics[width=8.5cm]{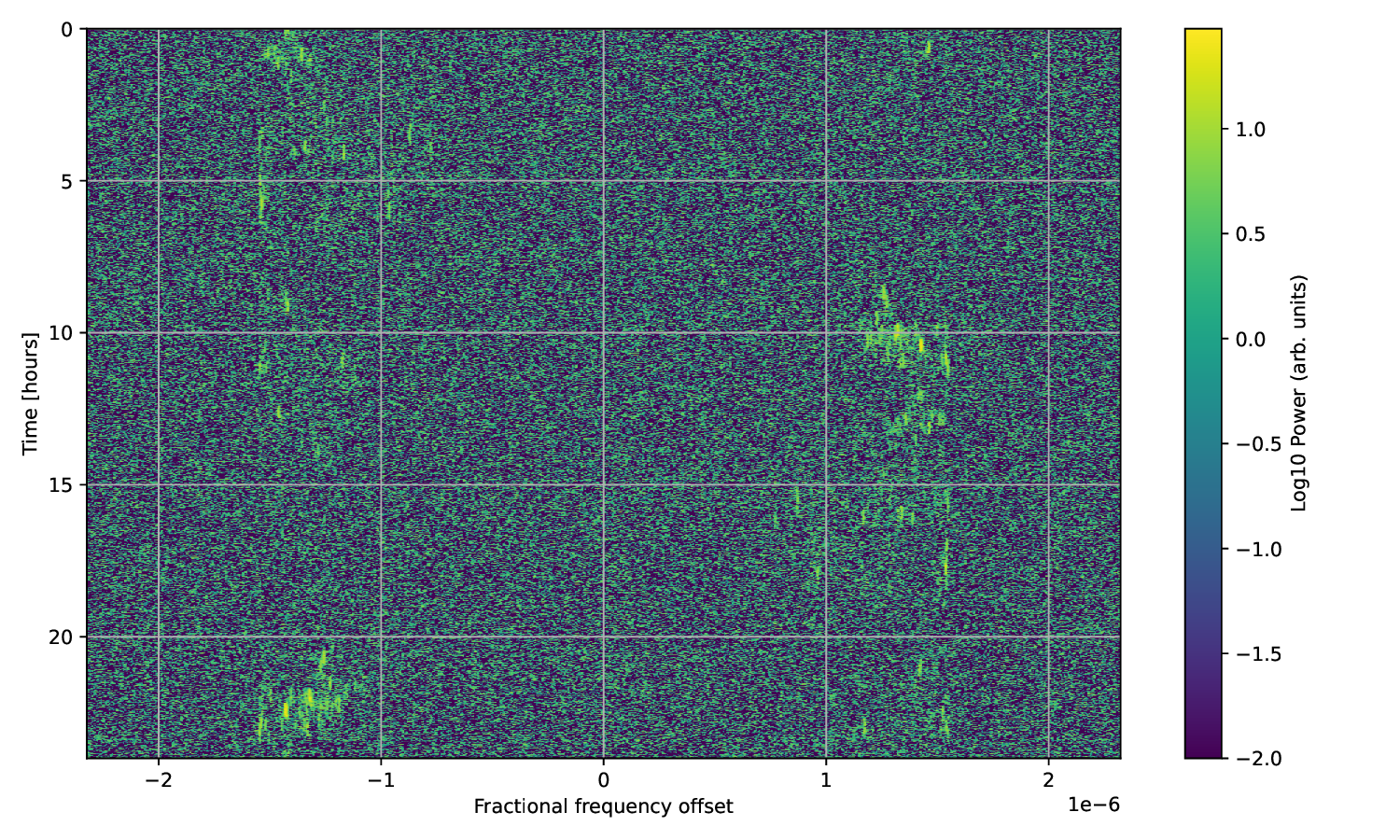} 
\includegraphics[width=8.5cm]{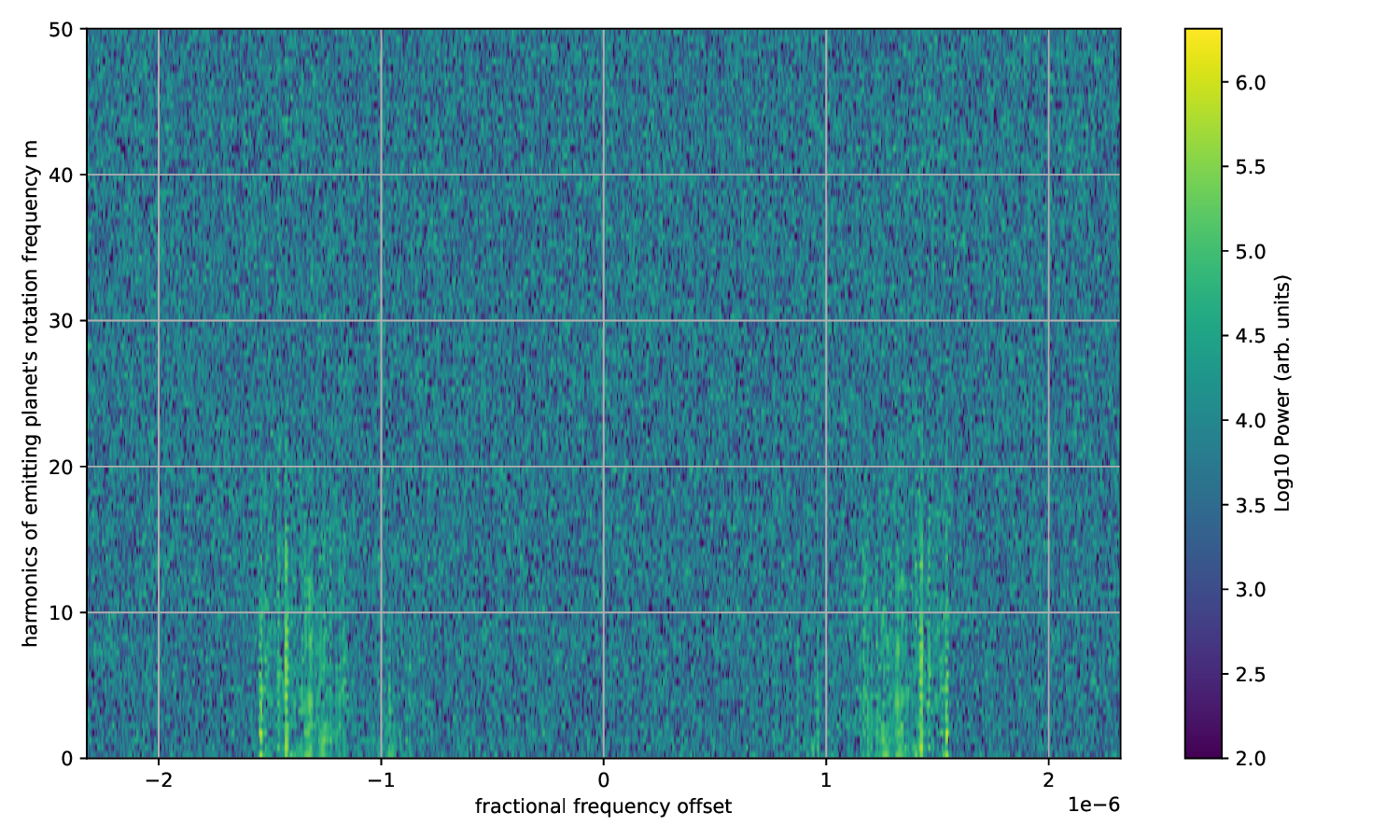}
\end{center}
\caption{Left: spectrogram with the signal of Fig. \ref{fig:spectrogram} and noise. Here, the noise level is set so that the brightest pixel has SNR=5. Right: the corresponding power spectra, $|\tilde{P(m \Omega, \Delta f)}|^2$.
\label{fig:noise}}
\end{figure*}

\begin{figure*}[t!]
\plotone{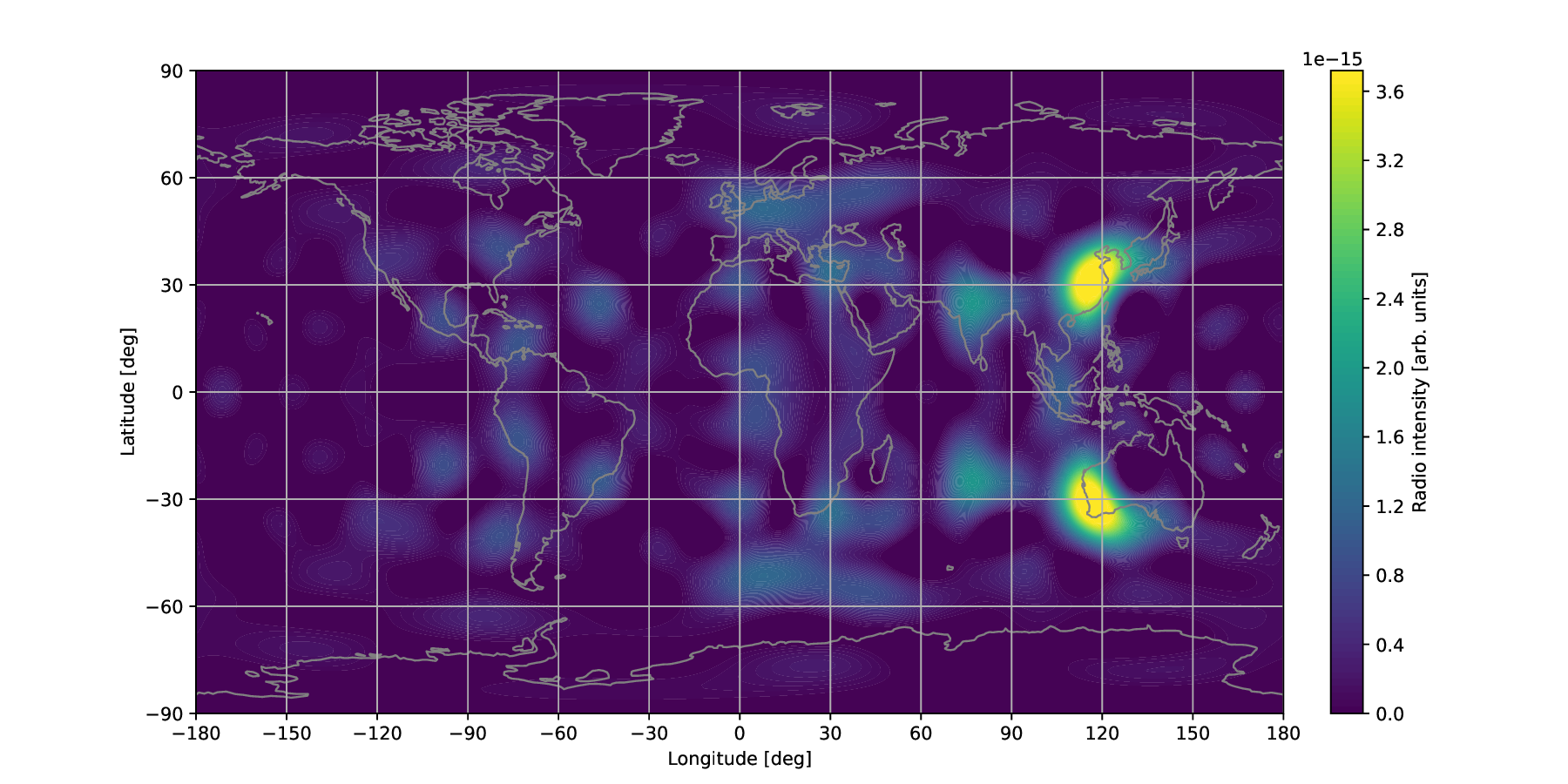} 
\caption{Reconstructed distribution of radio sources on Earth with noise (SNR of 5). Here Earth's rotation axis is perpendicular to the line of sight as in Fig. \ref{fig:spectrogram}. Comparing this reconstruction to Fig. \ref{fig:population} shows matching high-population regions. However, because identical signals from a given latitude could originate in either hemisphere, the reconstructed source map is symmetric about the equator, in other words, each actual transmitter appears twice, once in each hemisphere at the same latitude and longitude. In fact, most of the bright spots in the southern hemisphere originate from radio sources located in the northern hemisphere and do not actually exist. The regions that appear bright in the northern hemisphere are, from west to east, the U.S. West Coast, Mexico, Colombia and Venezuela, the U.S. East Coast, Western Europe, the area straddling the Sahara Desert around $20^\circ$ N latitude, West Asia, India, Indonesia, eastern China, and Japan; in the southern hemisphere, the only such region is Brazil.
\label{fig:map_noise}}
\end{figure*}

The corresponding spectrogram and power spectrum are shown in Fig.~\ref{fig:noise}. Fig. \ref{fig:map_noise} shows the reconstructed radio-source map obtained by extracting coefficients up to multipoles $\ell=20$ from noisy data. Despite the modest SNR, it recovers planet-scale structure. Comparison with Fig. \ref{fig:population} shows close correspondence with major population centers and continental patterns, with bright concentrations along the U.S. West and East Coasts and Mexico, Western Europe, West Asia-India, maritime Southeast Asia, eastern China, and Japan; in the southern hemisphere, Brazil stands out. Fainter sources remain buried in the noise at this SNR. Because rotational Doppler signatures are symmetric about the equator, the map is necessarily mirror-symmetric, and many apparent southern features are reflections of northern sources rather than distinct emitters. Even with this north-south ambiguity, the reconstruction robustly retrieves the gross distribution of radio activity, with continental landmasses and ocean basins reflected as, respectively, bright and dim regions. With higher SNR, the reconstruction would recover fainter cities and improve the spatial resolution of the inferred distribution.

%%%%%%%%%%%%%%%%%%%%%%%%%%%%%%%%%%%%%%%%%%%%%%%%%%%%%
\section{Discussion and Summary\label{sec:summary}}
%%%%%%%%%%%%%%%%%%%%%%%%%%%%%%%%%%%%%%%%%%%%%%%%%%%%%

Time-resolved narrowband spectra encode large-scale geography and activity on unresolved planets, moving beyond mere detection. Although a distant exoplanet appears as a point source in radio images, analysis of its time-varying narrowband signal can recover the spatial distribution of transmitters, which is likely to correlate with population centers, continental layouts, and climate zones. This enables a new class of technosignature measurements, from detection to cartography, by tracking rotationally modulated drifts, correcting long-term orbital motion, and coherently stacking many cycles to retrieve low-order spherical harmonics of the transmitter distribution, turning narrowband detections into quantitative constraints on planetary geography and activity. With precise observations, one could infer both the existence of an extraterrestrial civilization and key aspects of the civilization and host planet. This capability would mark a new stage for SETI, advancing from detection to mapping and characterization of a distant civilization's activity.

To demonstrate the method, we modeled Earth as an analog of an emitting technological planet hosting an extraterrestrial civilization and performed end-to-end simulations. We emphasize that the Earth-based population map used in our simulations is employed only as a convenient proxy for a generic “technosignature activity map.” We do not claim that present-day terrestrial emitters are both ultra-narrowband and distributed in proportion to population density. Instead, our goal is to demonstrate that, if an extraterrestrial civilization were to produce many narrowband (or narrow spectral-component) emitters whose spatial distribution traces centers of activity, then the rotational Doppler imprint could be used to reconstruct that distribution.

It is also useful to place our approach alongside two other ideas that have been discussed for extracting rotation-related information from technosignatures. One class of approaches focuses on the drift rate (and, for sufficiently long baselines, curvature) of a narrowband line, using the time evolution of its Doppler shift to constrain line-of-sight acceleration and thereby rotation and/or orbital motion. This is most naturally applicable when a single (or a few) persistent carriers can be tracked over extended time \citep{2019ApJ...884...14S,2022ApJ...938....1L}. Another class exploits intensity variations as a directive beam sweeps past the observer, where attenuation away from the beam center may encode a rotation or scan period when repeated beam crossings are observed and the beam pattern/pointing is sufficiently stable \citep{2010AsBio..10..475B,2010AsBio..10..491B,2020IJAsB..19..299G}. By contrast, our method is intended for an unresolved planet hosting many distributed narrowband emitters (a leakage-like ensemble), where near-horizon emission makes the approaching and receding limbs contribute prominently to the aggregate spectrum. Rather than relying on repeated passages through a narrow beam, our method uses the frequency-time structure imprinted by rotation to recover low-order information about the emitter distribution.

In this study, we assumed that all radio sources on the planet emit at the same frequency. This assumption, however, may be unrealistic in practice, when considering the situation on Earth. Nevertheless, it is conceivable that frequencies are standardized on a regional basis. Since the results presented here are expressed in terms of the ratio between the Doppler-induced frequency offset and the original frequency, observational data represented by this ratio would yield equivalent plots. Therefore, it is not necessary in practice for all radio sources to have identical frequencies.

Frequency agility and broadband modulation provide an additional limitation that depends on the variability timescale and span. If a transmitter’s center frequency does not change appreciably over the planetary rotation, our method applies directly. Many terrestrial emissions are effectively continuous (or have stable center frequencies) on hour timescales relevant to a planetary rotation period \citep{Sullivan,Saide}. If the center frequency hops or sweeps rapidly compared to the integration time of a few minutes, such as in frequency-hopping spread-spectrum links or frequency-agile radars that hop pulse-to-pulse or burst-to-burst (microseconds to milliseconds timescales), the emission is observed as an effectively broadened feature whose intrinsic fractional width reflects the hopping/sweep span. In this case, the rotational Doppler imprint remains recoverable provided this effective width is not much larger than the rotational Doppler span. In contrast, intermediate variability on minute-to-hour timescales can cause a narrowband feature to migrate across frequency bins between successive time steps, complicating coherent tracking and the present inversion. Terrestrial analogues include HF systems employing Automatic Link Establishment, where the operating channel can change on minute-to-tens-of-minutes timescales, and trunked land-mobile systems where channel assignments can change on tens-of-seconds to minutes timescales. Addressing this regime is left for future work.

Another major assumption of this study is that individual transmitters are extremely narrowband. Earth’s equatorial rotation speed is $470\,\mathrm{m\,s^{-1}}$, so the maximum fractional Doppler shift is $\sim 1.6 \times 10^{-6}$. Rotation-induced modulation becomes apparent only when a transmitter’s intrinsic fractional bandwidth is comparable to or smaller than this scale. On Earth, systems with fractional bandwidths as narrow as $10^{-6}$ are relatively uncommon. However, the effect would be amplified for civilizations on planets that rotate faster or have larger radii. Indeed, Earth’s spin in its early phase is thought to have been about four times faster than today. If a civilization arose somewhat earlier on a planet larger than Earth, the rotational Doppler signatures could be enhanced by several times relative to those modeled here.

The scenario considered in this work can be tested through Earth observations from the Moon. In this case, however, the Earth's rotation axis, as seen from the Moon, varies due to the Moon's orbital motion around the Earth: while $\theta=23.4^\circ$ remains fixed, $\phi$ varies at a constant rate with a period of $27.3$~days. Under such circumstances, the temporal variation of the spectrum would be more complex. Nevertheless, on timescales of a few days, the results should not differ significantly from those presented here. Similar verification is also possible via observations of Earth's moon-reflected radio emissions \citep{SullivanKnowles,McKinley}. Although the Moon's orbital motion changes the apparent orientation of Earth's rotation axis and thus complicates the spectral time series, the relative motion of Earth and Moon is well characterized and can be modeled and corrected for in such observations. Just as the first maps of our own world revolutionized exploration, the first maps of other technological worlds may reshape our understanding of our place in the cosmos.

%%%%%%%%%%%%%%%%%%%%%%%%%%%%%%%%%%%%%%%%%%%%%%%%%%%%%
\begin{acknowledgments}
The author would like to thank S. Yoshiura and Y. Fujii for valuable discussions. The author is partially supported by JSPS KAKENHI Grant Numbers 20H00180, 21H01130, 21H04467, 24H01813 and 25K21670 and Bilateral Joint Research Projects of JSPS 120237710.
\end{acknowledgments}

%%%%%%%%%%%%%%%%%%%%%%%%%%%%%%%%%%%%%%%%%%%%%%%%%%%%%
\appendix

\section{Reconstruction method formalism}
\label{app:reconstruction}

Here we summarize the details of the calculation for reconstructing the radio-source distribution in the delta-function model.

We consider an unresolved spherical planet of radius $R$ rotating rigidly at angular frequency $\Omega$, and the rotation-axis direction is specified in spherical coordinates ($\theta, \phi$). We adopt a planet-centered Cartesian frame in which the observer lies on the $+x$ axis and a surface element is labeled by latitude $\alpha$ and longitude $\delta$ (increasing eastward). At time $t$ its coordinate along the line of sight and its line-of-sight velocity are denoted by $x(\alpha,\delta,t)$ and $v_{\rm los}(\alpha,\delta,t)$, respectively. The rest-frame emission frequency is $f_0$ and the surface distribution of transmitters (or specific intensity) is denoted as $I(\mu, \delta)$, where $\mu \equiv \sin\alpha$. The observed spectrum is written as $P(t,\Delta f)$ as a function of time and frequency offset $\Delta f \equiv f-f_0$.

For simplicity, we take the rotation axis to be along the $z$ direction ($\theta=\phi=0$). In this case, we have,
\begin{equation}
\frac{x(\alpha, \delta, t)}{R} = \cos \alpha \cos(\Omega t + \delta), \qquad
v_{\mathrm{los}}(t)(\alpha, \delta, t) = - \Omega R \cos\alpha \sin(\Omega t + \delta).
\end{equation}
Here, we define $B \equiv f_0 R \Omega/c$, which is the maximum rotational Doppler frequency offset for an equatorial transmitter. Further, we approximate both the frequency characteristics and the beam pattern by delta functions. Noting $\cos\alpha = \sqrt{1-\mu^2}$, our emission model is,
\begin{equation}
P(t, \Delta f)
= \int_{-1}^1 d\mu \int_0^{2\pi} d\delta ~ I(\mu, \delta)
~ \delta(\Delta f + B \sqrt{1- \mu^2} \sin(\Omega t + \delta))
~ \delta(\sqrt{1- \mu^2} \cos(\Omega t + \delta)),
\end{equation}
where the first and second delta functions correspond to the frequency characteristics and the beam pattern, respectively. To perform the integration with respect to $\mu$ and $\delta$, we define the following functions from the arguments of the delta function.
\begin{eqnarray}
F(\mu, \delta) &\equiv& \Delta f + B \sqrt{1- \mu^2} \sin(\Omega t + \delta), \\
G(\mu, \delta) &\equiv& \sqrt{1- \mu^2} \cos(\Omega t + \delta).
\end{eqnarray}
These functions have zeros at the following values of $\mu$ and $\delta$:
\begin{equation}
\mu_0 = \pm \sqrt{1 - \left(\frac{\Delta f}{B}\right)^2}, \qquad
\delta_0 = -\Omega t - \frac{\pi}{2}\,\mathrm{sgn}(\Delta f) \quad (\mathrm{mod}\ 2\pi),
\end{equation}
where $\mathrm{sgn}(\Delta f)$ is the sign function of $\Delta f$ (positive for blueshift and negative for redshift). The Jacobian of these functions at $\mu = \mu_0$ and $\delta = \delta_0$ is evaluated as,
\begin{equation}
|J| \equiv
\left|\begin{array}{cc}
\frac{\partial F}{\partial \mu} & \frac{\partial F}{\partial \delta} \\
\frac{\partial G}{\partial \mu} & \frac{\partial G}{\partial \delta}
\end{array}\right|_{\mu=\mu_0, \delta=\delta_0}
= B \sqrt{1 - \left(\frac{\Delta f}{B}\right)^2}.
\end{equation}
Then, the integration with respect to $\mu$ and $\delta$ can be performed to obtain,
\begin{equation}
P(t, \Delta f) =
\frac{1}{B \sqrt{1 - (\Delta f/B)^2}}
\left[ 
I \left( \sqrt{1 - \left(\frac{\Delta f}{B}\right)^2}, -\Omega t - \frac{\pi}{2}\,\mathrm{sgn}(\Delta f) \right) 
+ I \left( -\sqrt{1 - \left(\frac{\Delta f}{B}\right)^2}, -\Omega t - \frac{\pi}{2}\,\mathrm{sgn}(\Delta f) \right)
\right].
\end{equation}

Next, we expand the distribution of radio sources, $I(\alpha, \delta)$, with spherical harmonics as,
\begin{equation}
I(\alpha, \delta) = \sum_\ell \sum_{m = - \ell}^{\ell} a_{\ell m} Y_{\ell m}(\pi/2 - \alpha, \delta).
\end{equation}
where,
\begin{equation}
Y_{\ell m}(\vartheta, \delta)
= \sqrt{\frac{2 \ell + 1}{4 \pi}} \sqrt{\frac{(\ell - m)!}{(\ell + m)!}} P^m_\ell (\cos\vartheta) e^{i m \delta}
\equiv A_{\ell m} P^m_\ell (\cos\vartheta) e^{i m \delta},
\end{equation}
and $P_\ell^m$ is the associated Legendre function. Then, $P(t, \Delta f)$ can be rewritten as,
\begin{equation}
P(t, \Delta f) =
\sum_\ell \sum_{m = -\ell}^{\ell} a_{\ell m} A_{\ell m} e^{-i m \Omega t} (- \mathrm{sgn}(\Delta f) i)^m
\frac{P_\ell^m(\sqrt{1- (\Delta f/B)^2}) + P_\ell^m(-\sqrt{1- (\Delta f/B)^2})}{B \sqrt{1- (\Delta f/B)^2}}
\end{equation}

We now take the Fourier transform with respect to $t$.
\begin{eqnarray}
\tilde{P}(\omega, \Delta f)
&=& \int_{-\infty}^\infty P(t, \Delta f) e^{i \omega t} dt \nonumber \\
&=& \sum_{\ell'} \sum_{m' = -\ell'}^{\ell'} a_{\ell' m'} A_{\ell' m'} \delta(\omega - m' \Omega) (- \mathrm{sgn}(\Delta f) i)^{m'}
\frac{P_{\ell'}^{m'}(\sqrt{1- (\Delta f/B)^2}) + P_{\ell'}^{m'}(-\sqrt{1- (\Delta f/B)^2})}{B \sqrt{1- (\Delta f/B)^2}}
\end{eqnarray}
Substituting $\omega = m\Omega$, the summation over $m'$ is performed to give,
\begin{equation}
\tilde{P}(m \Omega, \Delta f) =
\sum_{\ell' \geq |m|} a_{\ell' m} A_{\ell' m} (- \mathrm{sgn}(\Delta f) i)^m 
\frac{P_{\ell'}^m(\sqrt{1- (\Delta f/B)^2}) + P_{\ell'}^m(-\sqrt{1- (\Delta f/B)^2)}}{B \sqrt{1- (\Delta f/B)^2}}
\end{equation}
Furthermore, multiplying this by the associated Legendre functions and by $\Delta f$, and integrating the product over $\Delta f=0 \rightarrow B$,
\begin{eqnarray}
I_{\ell m}
&=& \int_0^B \tilde{P}(m \Omega, \Delta f)
    \left[ P_\ell^m(\sqrt{1- (\Delta f/B)^2}) + P_\ell^m(-\sqrt{1- (\Delta f/B)^2}) \right] \Delta f ~ d\Delta f \\
&=& \sum_{\ell' \geq |m|} a_{\ell' m} A_{\ell' m} (-i)^m \nonumber \\
&& \times \int_0^B  \frac{P_{\ell'}^m(\sqrt{1- (\Delta f/B)^2}) + P_{\ell'}^m(-\sqrt{1- (\Delta f/B)^2})}{B \sqrt{1- (\Delta f/B)^2}} \nonumber \\
&& \times \left[ P_\ell^m(\sqrt{1- (\Delta f/B)^2}) + P_\ell^m(-\sqrt{1- (\Delta f/B)^2}) \right] \Delta f ~ d\Delta f
\end{eqnarray}
Here, changing variables from $\Delta f$ to $x=\sqrt{1-(\Delta f/B)^2}$, we have
\begin{eqnarray}
I_{\ell m}
&=& \sum_{\ell' \geq |m|} a_{\ell' m} A_{\ell' m} (-i)^m B  \int_0^1  \left[ P_{\ell'}^m(x) + P_{\ell'}^m(-x) \right] \left[ P_\ell^m(x) + P_\ell^m(-x) \right] dx \\
&=& \sum_{\ell' \geq |m|} a_{\ell' m} A_{\ell' m} (-i)^m B \nonumber \\
&& \times \int_0^1 \left[ P_{\ell'}^m(x) P_\ell^m(x) + P_{\ell'}^m(-x) P_\ell^m(-x) + P_{\ell'}^m(x) P_\ell^m(-x) + P_{\ell'}^m(-x) P_\ell^m(x) \right] dx \nonumber \\
&=& \sum_{\ell' \geq |m|} a_{\ell' m} A_{\ell' m} (-i)^m B
    \int_0^1 \left(1+(-1)^{\ell+m}\right) \left[ P_{\ell'}^m(x) P_\ell^m(x) + P_{\ell'}^m(-x) P_\ell^m(-x) \right] dx \nonumber \\
&=& \sum_{\ell' \geq |m|} a_{\ell' m} A_{\ell' m} (-i)^m B \left(1+(-1)^{\ell+m}\right)
    \int_{-1}^1 P_{\ell'}^m(x) P_\ell^m(x) dx \nonumber \\
&=& \sum_{\ell' \geq |m|} a_{\ell' m} A_{\ell' m} (-i)^m B \left(1+(-1)^{\ell+m}\right)
    \frac{2 (\ell + m)!}{(2 \ell + 1) (\ell - m)!} \delta_{\ell \ell'} \nonumber \\
&=& a_{\ell m} A_{\ell m} (-i)^m B \left(1+(-1)^{\ell+m}\right) \frac{2 (\ell + m)!}{(2 \ell + 1) (\ell - m)!}
\end{eqnarray}
where we used the parity and orthogonality of the associated Legendre function,
\begin{eqnarray}
&& P_{\ell}^m(-x) = (-1)^{\ell + m} P_{\ell}^m(x), \\
&& \int_{-1}^1 P^m_\ell(x) P^{m}_{\ell'}(x) dx = \frac{2 (\ell + m)!}{(2 \ell + 1) (\ell - m)!} \delta_{\ell \ell'}.
\end{eqnarray}
We see that, if $\ell+m$ is odd, $I_{\ell m}=0$. On the other hand, if $\ell+m$ is even, the coefficients in the spherical-harmonic expansion are given by
\begin{equation}
a_{\ell m}
= \frac{i^m}{2 B} \sqrt{\pi (2 \ell + 1)} \sqrt{\frac{(\ell - m)!}{(\ell + m)!}} I_{\ell m},
\end{equation}
which coincides with Eq.~(\ref{eq:alm}). Therefore, the radio-source distribution can be partially reconstructed by $a_{\ell m}$ with even $\ell+m$. For general frequency characteristics and beam patterns, including those with Gaussian profiles, it is difficult to perform analytical calculations as described above; therefore, numerical computation would be required. While the demonstration presented here assumes a simplified case involving a delta-function profile, it remains highly useful for intuitively understanding how planetary radio source maps are reconstructed from observed spectrograms.

\bibliography{SETI_ver4}{}
\bibliographystyle{aasjournalv7}

%% Include this line if you are using the \added, \replaced, \deleted
%% commands to see a summary list of all changes at the end of the article.
%\listofchanges

\end{document}